\documentstyle[aps,prb,eqsecnum,amsbsy,epsf]{revtex}
\author{Kristian K.  M\"uller-Nedebock$^{*}$, Thomas A. Vilgis}
\address{Max-Planck-Institut f\"ur  Polymerforschung,
Postfach 3148, 55021 Mainz, Germany}
\title{Dynamics of dense Polyelectrolyte Solutions}

\date{1998 }

\begin{document}

\maketitle

\begin{abstract}
We investigate a system of dense polyelectrolytes in solution.  The
Langevin dynamics of the system with linearized hydrodynamics is formulated
in the functional integral formalism and a transformation made to
 collective coordinates.  Within a dynamical Random Phase
Approximation (RPA) integration over the counter- and salt ions produces the
Debye-H\"uckel-like  screening of the Coulomb interactions
with dependence on the frequency only as part of a more complicated coupling
structure.  
We investigate the dynamics of the structure
factor as well as the collective diffusion coefficient and comment upon
the viscosity of the 
whole system of polymers with counterions and fluid in the simplest
approximation.  
The coupling of the
various components of the system produces nontrivial diffusive behavior.
We draw conclusions about the relationship of the three length
scales in the present system, { i.e.} the static screening length, the 
hydrodynamic screening length and the Debye length.  
\end{abstract}


\section{Introduction}

The statistical mechanics of polyelectrolytes has recently been the object of
much research, of both
analytical nature  and simulations, 
ranging from a description of the flexible single-chain
properties~\cite{StapperLiverpool,BarratJoanny,LiWitten,Khokhlov,Stevens,Micka1,Micka3} 
to the properties of the dense melts in collective
coordinates~(refs 8-10
and references therein).  
The presence of the long-ranged Coulomb interactions makes such systems
completely different to uncharged ones.  
Underlying many of these approaches is the electrostatic screened interaction
according to 
Debye-H\"uckel, of the form 
\begin{equation}
V_{\rm DH} ({\bf r}) = k_{B}T\frac{\lambda_{B}}{|{\bf r}|} e^{-\kappa |{\bf r}|}.
\end{equation}
Here $\kappa$ represents the screening parameter, or its inverse the
Debye-H\"uckel screening length, and $\lambda_{B}$ is called the Bjerrum
length,
\begin{equation}
\lambda_{B} = \frac{e^{2}}{4\pi \epsilon k_{B}T}.
\end{equation}
Here $\epsilon$ is the dielectric constant of the fluid, and $k_{B}T$ the
temperature multiplied by the Boltzmann constant.
The parameter, $\kappa$, is proportional to the Bjerrum length and to the
concentration of the counterions.  Its inverse is the Debye screening length.
The effective potential  
can be derived from the linearized Poisson-Boltzmann
equation\cite{PoissonBoltzmann} 
or by utilizing
collective density variables together with the random phase approximation in
statistical mechanics
(RPA) approach\cite{VilgisBorsali}
and integration over salt and counterion degrees of
freedom, which procedure again reproduces that form of effective screened interaction.
In other models the mean spherical approximation (MSA) is used.\cite{MSA}
However, 
whether the Yukawa form of the pairwise potential is truly accurate for
macromolecular systems still remains a debatable question and even for the
understanding of the critical behavior in the case of the simple electrolyte a
debate concerning an appropriate minimal model exists (See, e.g. Levin and
Fisher\cite{LevinFisher}). 

On the statistical mechanical level, the properties of the single chain have
been under much discussion as to the precise dependence upon $\kappa$ of the
properties of the chain, which is expected to behave as with an enhanced
persistence length due to the electrostatic interaction.  Recent
approaches to the problem are a treatment by Barrat and
Joanny\cite{BarratJoanny,BarratJoannyRev} and a renormalization treatment by Liverpool and
Stapper.\cite{StapperLiverpool}  On the level of systems also of higher than
extremely dilute concentrations recent work by Muthukumar\cite{Muthukumar}
has been based upon the
assumption of an electrostatically altered effective, Gaussian step-length.
As regards the implementation of the RPA
it has also been shown\cite{VilgisBorsali} that this approximation
produces qualitatively correct results.  It needs to be altered when strong,
attractive interactions can cause complexation\cite{VilgisHaronska} or when
the effects of the strength of the potential are such that the Gaussian
assumption must fail.

The dynamics of polyelectrolytes has recently been addressed in
scaling-theoretic approach by Dobrynin {\em et al}.\cite{Dobrynin1,Dobrynin2}
An important experimental result in those dynamical considerations is the
Fuoss law\cite{Fuoss}
which states the proportionality of
the viscosity and the inverse square root of the polyion concentration at
sufficiently high densities.  This was recently confirmed by Dobrynin, Rubinstein and 
Colby.\cite{Dobrynin1,Dobrynin2} Other treatments of this law have been by
Witten and Pincus\cite{WittenPincus} and by Rabin.\cite{Rabin} 
Previously, a theoretical treatment\cite{VB1,VB2} of the
polyelectrolyte solution viscosity has been given by one of the authors by
making use of the results of a mode-mode coupling approximation by Hess and
Klein\cite{HessKlein} in which a peak of the viscosity was derived.
Recent experimental work by Isambert, Ajdari, Mitnuk, Viovy and 
Prost\cite{IsambertEtAl} has
highlighted a physical instability of a polyelectrolyte solution occurring
when this solution is subjected to an external electric field.

In this paper we formulate the dynamics of a system of polyions with salt and
counterions in the Langevin approach, and couple the dynamics of these
objects to a solvent which is assumed to be governed by
linearized hydrodynamics.  
It is possible to bring the
different length scales inherent to this problem, {i.e.} the charge screening
length, the hydrodynamic screening and the polymer length,
into relation with each other.  It is assumed  that the
system is sufficiently dense and the chain conformations close to the Gaussian
ideal to permit the use of a random phase approximation
(RPA) in the context of the dynamics (see, e.g. ref 25).
Furthermore, it
is assumed that 
linearized hydrodynamic properties of the solvent apply and that the
investigation is limited to the semi-dilute regime in which the 
effects of entanglements of the macromolecular strands can be
neglected. Importantly we find that this coupling of the dynamics produces
results which are at variance with calculations where an effective
Debye-screened potential were used straight away.

The paper is outlined as follows.  Section~II introduces the
dynamical equations of the model system.   These expressions are reformulated
in terms of a set of collective variables and expressed through the formalism
of Martin, Siggia and Rose\cite{MSR} and of Peliti and De
Dominicis.\cite{PelitiDominicis}   The subsequent sections and subsections under
III  show the results for
computations of the correlation functions: this is the
dynamical structure factor from which we draw conclusions about the
diffusional behavior. The final section recounts the results, discusses the
hydrodynamic aspects, and
concludes on the interplay of the screening lengths.


\section{The Model}

\subsection{The Langevin Equations}

The model consists of the 
Edwards Hamiltonian for the flexible polyions with the
excluded volume interaction written up to second virial order with the
electrostatic potential containing the appropriate Bjerrum length,
$\lambda_B = e^2/(4\pi\epsilon k_BT)$, for the
solvent.  The dielectric constant for the solvent is given by $\epsilon =
\epsilon_{r} \epsilon_{0}$, $k_{B}T$ the Boltzmann constant multiplied by
the temperature, and $e$ is the electronic charge  (For the sake of simplicity
only monovalent systems
are discussed, although changing valences would present no significant problem
in the formalism as used and approximated here.).
Furthermore, counterions and salt ions occupy the solvent such that
overall electrical neutrality is maintained.
We label the four different species of molecule, i.e. polyion, counterion, and
the two types of salt ion, by Greek subscripts $\alpha=1,\ldots,4$.  There are
exactly $N_{\alpha}$ identical ions of each species, with $N_{3}=N_{4}$, which
have position coordinates given by ${\bf r}_{p \alpha}(s,t)$, where the contour
length $s$ is only
a relevant variable for the macroions ($\alpha=1$).  Apart from this no further
internal degrees of freedom are associated with any of the system
constituents.   The complete Hamiltonian expresses the flexibility of the
chains, the excluded volume contributions and the (unscreened) Coulomb
interaction.  
\begin{eqnarray}
H & = & \frac{3}{2\ell}\sum_{p=1}^{N_1} \int_{0}^{L_{\alpha}} 
ds\, \left( \frac{\partial {\bf r}_{p1}(s,t)}{\partial
s} \right)^2
+
\frac{1}{2}\sum_{\alpha=1}^{4}\sum_{\beta=1}^{4}\sum_{p_{\alpha}=1}^{N_{\alpha}}
\sum_{q_{\beta}=1}^{N_{\beta}}
\int_{0}^{{L_{\alpha}}}\!\int_{0}^{{L_{\beta}}}
ds_{p_{\alpha}}\,ds_{q_{\beta}}\, \times \nonumber \\
& & \times \left( 
w_{\alpha\beta}\delta\left( {\bf r}_{p\alpha}(s_{i_{\alpha}},t)-{\bf r}_{q\beta}(s_{j_{\beta}},t)\right) +
M_{\alpha\beta} \frac{\left( 1 - \delta_{\alpha\beta}
\delta_{i_{\alpha}j_{\beta}}(1-\delta_{1i_{\alpha}})\right)}{\left| 
{\bf r}_{p\alpha}(s_{i_{\alpha}},t)-
{\bf r}_{q\beta}(s_{j_{\beta}},t)\right|}\right)
\end{eqnarray}
The Kuhn length of the chains is given by $\ell$, and the numerator 
which contains Kronecker delta functions of the
term expressing the Coulomb interaction is a technical description to exclude
self-interaction by the point particles.
Formally, in order for the general form above to make sense, the lengths,
$L_{\alpha}$, of the counterions and salt ions are one monomer.
The Coulomb interaction is represented by the matrix $M_{\alpha\beta}$ of the
appropriate interaction between the polymer and the counterions of the
system:
\begin{equation}
M_{\alpha\beta} =  \lambda_B k_B T \left( \begin{array}{cccc}
f^2 & - f & +f & -f \\
-f & + 1 & -1 & +1 \\
+f & - 1 & +1 & -1 \\
-f & +1 & -1 & + 1
\end{array} \right) .
\end{equation}
The condition of electrical neutrality implies that $
N_1 L f - N_2 + N_3 - N_4 = 0$ must hold.
The quantity $f$ represents the constant, linear charge density of the polymer.  The
matrix $w_{\alpha\beta}$ represents the excluded volume interaction between the
various components of the system.
The Hamiltonian and the hydrodynamics then lead to the
coupled, stochastic Langevin equations
of the system for the chain molecules, the counterions, and the fluid
velocity field.\cite{OonoFreed}  When stick boundary conditions are employed the stochastic
equations can be written as
\begin{eqnarray}
{\bf L}_p^\alpha & \equiv & - \frac{\partial}{\partial t} {\bf r}_{p\alpha}(t)
+ {\bf v}[{\bf r}_{p\alpha}(t)] - \frac{1}{\zeta_\alpha} \frac{\delta
H}{\delta {\bf r}_{p\alpha}(t)} + {\bf f}_{p\alpha}(t) = 0 \\
{\bf L}^v & \equiv & - \frac{\partial}{\partial t} {\bf v}({\bf x},t)
+ \left\{ \eta\nabla^2{\bf v}({\bf x},t) - \sum_{p=1}^{N_1} \int_0^{L_{1}} ds\,
\frac{\delta H}{\delta {\bf r}_{p1}(s,t)}\delta\left({\bf x}-{\bf
    r}_{p1}(s,t)\right) 
\right.\nonumber \\
& & - \left. \sum_{\alpha=2}^4\sum_{p=1}^{N_\alpha} \frac{\delta H}{\delta
{\bf r}_{p\alpha}(t)} \delta\left({\bf x} - {\bf r}_{p\alpha}(t)\right) +
{\bf f}_v({\bf x},t) \right\}_\perp = 0 .
\end{eqnarray}
The random forces, ${\bf f}_{p\alpha}(s_{p\alpha},t)$ and ${\bf f}_{v}({\bf x},t)$
for the dynamics of the molecules and the fluid dynamics, respectively,
are correlated as Gaussians and
satisfy
the fluctuation--dissipation theorem given by 
$
\left\langle {\bf f}_{p\alpha}(s_{p_{\alpha}},t){\bf f}_{q\beta}(s_{q_{\beta}},t') \right\rangle
= $ $ \frac{2}{\zeta_{\alpha}}\delta_{\alpha\beta}{\bf 1}\delta_{p_{\alpha}q_{\beta}}
\delta (t-t') \delta(s_{p_{\alpha}}- s_{q_{\beta}})$ and  $
\left\langle {\bf f}_{v}({\bf x}, t){\bf f}_{v}({\bf x}',t') \right\rangle  = $ $ - 2 \eta {\bf
  1} \nabla^{2}_{\bf x}\delta ({\bf x}- {\bf x}') \delta(t-t')$.
We are also working in the
case of the incompressible fluid continuum\cite{OonoFreed} and,
hence, impose the transverse condition upon the velocity field of the fluid
of bare viscosity $\eta$.  The components have friction coefficients
$\zeta_{\alpha}$ which are in general different.  
The energy units above have been chosen such that $k_{B}T = 1$ and for
subsequent sections the distance units of the polymer will also be chosen such
that the prefactor $3/2\ell$ in the Hamiltonian vanishes.

        
\subsection{Functional Integral Formalism}

We now treat the set of coupled differential equations by using the functional
integral approach of Martin, Siggia and Rose\cite{MSR} and follow the
scheme of Fredrickson and Helfand\cite{FredricksonHelfand} used for uncharged
solutions of monodisperse polymers.  Here we shall be concerned primarily
with the collective behavior of the system dynamics, and, in following the
work of Fredrickson and Helfand, we shall later ignore contributions
in the collective variables which are of order higher than two.

By making use of causality and that the Jacobian of the transformation then
becomes unity\cite{Jensen}
the functional integral expressions become:
\begin{eqnarray}
Z  & =  & \left\langle \int \left\{
\prod_{\alpha=1}^{4}\prod_{p=1}^{N_{\alpha}} {\cal D} {\bf r}_{p\alpha}(t) {\cal D} {\bf
\hat{r}}_{p\alpha}(t)  \right\}
{\cal D}{\bf v}({\bf x},t){\cal D}{\bf \hat{v}}({\bf x},t)
\right. \nonumber \\ & & \left. \exp \left\{ +i\left[  \sum_{\alpha=1}^{4} 
\sum_{p=1}^{N_{\alpha}}\int_{-\infty}^{+\infty} dt \int_{0}^{L_{\alpha}}ds \, {\bf
\hat{r}}_{p\alpha}(t)\cdot{\bf L}_{p}^{\alpha} + \int_{-\infty}^{+\infty} dt
\int_{\Omega}d^3x \,
{\bf
{\hat{v}}}({\bf x},t)\cdot{\bf L}^v \right] \right\} \right\rangle .
\end{eqnarray}
Angular brackets indicate the averaging over Gaussian noise and $\Omega$
represents the volume.
Following ref~25 we introduce
collective variables in terms of collective density and response functions for each
species $\alpha$ (with a number,  $N_{\alpha}$ of molecules):
\begin{eqnarray} 
\rho_{\alpha}({\bf k},t)& = & \sum_{p=1}^{N_{\alpha}}
\int_{0}^{L_{\alpha}} ds \,\exp \left({+i{\bf k}\cdot {\bf
    r}_{p\alpha}(s,t)}\right) \\
{\boldsymbol{\sigma}}_{\alpha}({\bf k},t)&=& \sum_{p=1}^{N_{\alpha}}\int_{0}^{L_{\alpha}}ds\,  
  {\bf \hat{r}}_{p\alpha}(s,t) \,\exp \left({+i{\bf k}\cdot {\bf r}_{p\alpha}(s,t)}
  \right) \\
{\boldsymbol{\pi}}_{\alpha}({\bf k},t)& = & \delta_{\alpha 1}
({\bf 1}-{\bf \hat{k}\hat{k}})\cdot\sum_{p=1}^{N_{1}}\int _{0}^{L_{\alpha}}ds\,
\left( \frac{\partial^{2}{\bf r}_{1}(s,t)}{\partial s^{2}} \right) \, 
\exp\left({+i{\bf k}\cdot {\bf
    r}_{p1}(s,t)}\right) .
\end{eqnarray}
Clearly, the ${\boldsymbol{\pi}}_{\alpha}$ has no meaning when it relates to
point-like salt or counterions in the system.

If one were to impose the additional condition that the whole system finds
itself in a spatially varying, external field, which oscillates at a
frequency ${\tilde{\omega}}$, such as $2 {\tilde{V}}({\bf k})\cos {\tilde{\omega}} t$ 
the Langevin equations in the MSR formalism of
equation~2.5 would have an additional contribution in the exponent of:
\begin{equation}
{\cal L}_{\rm ext} = \sum_{\alpha}\sum_{\omega {\bf k}} 
 \frac{i}{\zeta_{\alpha}}  {\tilde{V}}({\bf k})\,{\bf k}\cdot\left( {\boldsymbol{\sigma}}_{\alpha}
({\bf k},\omega + {\tilde{\omega}}) +
{\boldsymbol{\sigma}}_{\alpha}({\bf k}, \omega - {\tilde{\omega}})\right) .
\end{equation}

The linearized hydrodynamic
formulation used here permits the integration over the fluid velocity fields
to be performed exactly and enables the determination of the
frequency--dependent structure factor of the molecular inclusions from which
the diffusion coefficient of these components is derived.
Consequently, one can write
\begin{equation}
Z = \int \left\{ \prod 
{\cal D} {\bf {r}} {\cal D}{\bf \hat{r}}\right\} \exp {-{\cal L}} ,
\end{equation}
where 
${\cal L} = {\cal L}_0 + {\cal L}_{\rm I} +{\cal L}_3$.
The variables and integrals are all expressed in terms of their of
Fourier representation, the random forces have been integrated and the
functional 
integral over the velocity fields, which occur only to quadratic order,
have been performed.
The functions above are given by
\begin{eqnarray}
{\cal L}_0 & = & -i \int_\omega \left\{ \sum_{p=1}^{N_1} \int_0^{L_{1}} ds\, {\bf
\hat{R}}_{p1}(s,\omega) \cdot \left[ i\omega {\bf R}_{p1}(s,\omega) +
\frac{1}{\zeta_1}\frac{\partial^2}{\partial s^2} {\bf R}_{p1}(s,\omega)
+ \frac{i}{\zeta_1} {\bf \hat{R}}_{p1}(s,\omega) \right] \right. \nonumber \\
& & \left.
+ \sum_{\alpha=2}^4\sum_{p=1}^{N_\alpha} {\bf \hat{r}}_{p\alpha}(\omega)
\cdot \left[ i \omega {\bf r}_{p\alpha}(\omega) + \frac{i}{\zeta_\alpha} {\bf
\hat{r}}_{p\alpha}(\omega) \right] \right\}
\\
{\cal L}_{\rm I} & = & i \sum_{\alpha=1}^4 \sum_{\beta=1}^{4}
\int_{\bf k}\int_\omega \frac{1}{\zeta_\beta}
\rho_\alpha
(-{\bf k},-\omega) \left[ 4\pi k^{-2} M_{\alpha\beta} + w_{\alpha\beta} \right]
i{\bf k}\cdot{\boldsymbol{\sigma}}_\beta({\bf k},\omega) \nonumber \\
& & + \sum_{\alpha=1}^4 \sum_{\beta=1}^{4}\int_{\bf k}\int_\omega ({\bf 1} - {\bf
\hat{k}\hat{k}})\cdot{\boldsymbol{\sigma}}_\alpha(-{\bf k},-\omega)\cdot \nonumber
\\ & & \cdot \left[ \frac{\eta k^2}{\omega^2+\eta^2k^4} ({\bf 1} - {\bf
\hat{k}\hat{k}})\cdot {\boldsymbol{\sigma}}_\beta({\bf k},\omega) +
\frac{1}{\omega + i\eta k^2} {\boldsymbol{\pi}}_\beta({\bf k}, \omega) \right] \\
{\cal L}_3 & = & i \sum_{\alpha=1}^{4}\sum_{\beta=1}^{4}\sum_{\gamma=1}^{4}
\int_{{\bf k}{\bf k}'}\int_{\omega\omega'}
\frac{1}{\omega + i\eta k^2}({\bf 1} - {\bf \hat{k}\hat{k}})\cdot
{\boldsymbol{\sigma}}_\beta(-{\bf k},-\omega)\cdot{\bf k}'\rho_\beta({\bf k}',\omega')
\nonumber \\
& & \times \left[
M_{\beta\gamma}({\bf k}')^{-2} + w_{\beta\gamma} \right] \rho_\gamma
({\bf k}-{\bf k}',\omega -\omega')
\end{eqnarray}
The friction coefficients for the polyion chain segments and the
counterions are given by $\zeta_1$ and $\zeta_2$, respectively.  The second
term of ${\cal L}_{I}$ and the contribution ${\cal L}_{3}$  come from the
coupling to the hydrodynamic medium with sticking boundary conditions.

Since we assume to be dealing with sufficiently dense systems below the onset
of entanglement effects, we can make use of the approximation that the chains
behave as Gaussians.  The random phase approximation is implemented utilizing
$\exp -{\cal L}_{0}$ as the statistical measure for the collective variables as
done by Fredrickson and Helfand.\cite{FredricksonHelfand}  The results are
summarized in the appendix.  Importantly, in the quadratic approximation
the collective variables $\rho_{\alpha}$ and its dynamical conjugate field ${\bf k}\cdot
{\boldsymbol{\sigma}}_{\alpha}$ decouple from $({\bf 1}-{\bf
  {\hat{k}\hat{k}}})\cdot {\boldsymbol{\sigma}}_{\alpha}$ and 
$({\bf 1}-{\bf
  {\hat{k}\hat{k}}})\cdot {\boldsymbol{\pi}}_{\alpha}$.  This decoupling and
the quadratic distribution also cause the cubic contribution ${\cal L}_{3}$ to
be a negligible contribution such that the hydrodynamics will not have any
effect upon $\rho_{\alpha}$ and  ${\bf k}\cdot
{\boldsymbol{\sigma}}_{\alpha}$ when the interactions are considered at this
order.

For the sake of simplicity, we perform the calculation for the system consisting of
polyions and the counterions leaving the component labels $\alpha =1$ and
$\alpha=2$ with concentrations given by $\rho_{1}$ and $\rho_{2}$,
respectively.  Extending the results for salt ions is straightforward.
Investigating the properties of the structure factor
requires the use of the collective variables which can be contained in a
supervector,
\begin{equation}
{\underline \psi} ({\bf k}, \omega) = \left( \rho_{1}({\bf k},\omega), \,\,\, {\bf
  k}\cdot {\boldsymbol{\sigma}}_{1}({\bf k},\omega),\,\,\, \rho_{2}({\bf
  k},\omega), 
  \,\,\,{\bf
  k}\cdot {\boldsymbol{\sigma}}_{2}({\bf k},\omega) \right) .
\end{equation}
Averages concerning these quantities can now be expressed as follows:
\begin{equation}
\left\langle {\sf A} \right\rangle_{\rm RPA} = {\cal N} \int {\cal D}{\underline \psi} \,
{\sf A} \,\exp -\left[ {\underline \psi}^{T}(-{\bf k}, \omega) \cdot  {\bf T}
\cdot {\underline \psi}({\bf k}, \omega)\right],
\end{equation}
Where the matrix ${\bf T}$ contains the contributions from the RPA and the
interaction term, ${\cal L}_{I}$.  An explicit form of this matrix is given
in the following section.


\section{The Dynamic Structure Factor}

After including the interactions and the two coupled density fields which are
not in the transverse direction, the computation of the structure factor
reduces to the identification of the appropriate elements of the inverse of
the matrix ${\bf T}$ given below.  The excluded volume interaction is 
assumed only between polyion chain segments,  such that we take the matrix
$w_{\alpha\beta}$
to be given by the expression $w_{\alpha\beta}= \delta_{1\alpha}\delta_{1\beta}v$.
Here we consider the case of 
large length scales and large times.  These conditions,
$k R_{g} \ll  1$ and $\omega \tau_{r}\ll  1$,  where in the units which have
been chosen $R_{g} = L/2$, and the Rouse time, $\tau_{r}=L^{2}\zeta_{1}\pi^{-2}$,
lead to the matrix in the quadratic supervector formulation,
\begin{equation}
{\bf T} =  \left(
{\begin{array}{cccc}
0 & {\displaystyle \frac {{i}}{{{ L\rho}_{1}}}} + 
{\displaystyle \frac {{{ \zeta}_{1}}\,{\omega}}{{{ \rho}_{1}}\,{k}^{2}
}} + {i}{v} + {\displaystyle \frac {{i}{ \lambda_{B}}{f}^{2}}{{
k}^{2}}} & 0 &  - {\displaystyle \frac {{i}{ \lambda_{B}}
{f}}{{k}^{2}}} \\ [2ex]
{\displaystyle \frac {{i}}{{{ L \rho}_{1}}}} - {\displaystyle 
\frac {{{ \zeta}_{1}}{\omega}}{{{ \rho}_{1}}{k}^{2}}} + {i}{v}
 + {\displaystyle \frac {{i}{ \lambda_{B}}{f}^{2}}{{k}^{2}}} & 
{\displaystyle \frac {\zeta_{1}}{{k}^{2}{{ \rho}_{1}}}} &  - 
{\displaystyle \frac {{i}{ \lambda_{B}}{f}}{{k}^{2}}} & 0 \\
 [2ex]
0 &  -{\displaystyle \frac {{i}\,{ \lambda_{B}}\,{f}}{{k}^{2}}}
& 0 & {\displaystyle \frac {{i}}{{{ \rho}_{2}}}} + 
{\displaystyle \frac {{{ \zeta}_{2}}{\omega}}{{{ \rho}_{2}}{k}^{2}
}} + {\displaystyle \frac {{i}{ \lambda_{B}}}{{k}^{2}}} \\ [2ex]
 - {\displaystyle \frac {{i}{ \lambda_{B}}\,{f}}{{k}^{2}}} & 0
& {\displaystyle \frac {{i}}{{{ \rho}_{2}}}} - 
{\displaystyle \frac {{{ \zeta}_{2}}{\omega}}{{{ \rho}_{2}}{k}^{2}
}} + {\displaystyle \frac {{i}\,{ \lambda_{B}}}{{k}^{2}}} &
{\displaystyle \frac {\zeta_{2}}{{k}^{2}{{ \rho}_{2}}}}
\end{array}}
\right).
\end{equation}
We shall be interested primarily in the diffusional behavior contained within
this matrix.  This requires only the consideration of the limits given above.

\subsection{Expression for the structure factor}

As a consequence we can write the polymer-polymer structure factor for the
case $kR_{g} \ll 1$.  By recalling the definition of the matrix ${\bf T}$, this
must be the 1,1 element of the inverse, ${\bf T}^{{-1}}$,
\begin{eqnarray}
G_{11} ({\bf k}, \omega) &  = &  
2\,{k}^{2}\rho_{{1}}\left (\zeta_{{1}}{k}^{4}+2\zeta_{{1}}{k}^{2}\lambda_{B}
\rho_{{2}}+\zeta_{{1}}{\zeta_{{2}}}^{2}{\omega}^{2}+\zeta_{{1}}{\lambda_{B}}
^{2}{\rho_{{2}}}^{2}+\rho_{{1}}{\lambda_{B}}^{2}{f}^{2}\zeta_{{2}}\rho_{{2}}
\right ){L}^{2}  \nonumber \\
& / &  \left( 2\,{k}^{6}\lambda_{B}\,{f}^{2}\rho_{{1}}L+{k}^{8}+2
L{k}^{8}v\rho_{{1}}+{
\lambda_{B}}^{2}{f}^{4}{\rho_{{1}}}^{2}{k}^{4}{L}^{2}+{\zeta_{{1}}}^{2}{\omega
}^{2}{k}^{4}{L}^{2}+{\zeta_{{1}}}^{2}{\omega}^{4}{\zeta_{{2}}}^{2}{L}^{2}
\right. \nonumber \\ && +
{\lambda_{B}}^{2}{f}^{4}{\rho_{{1}}}^{2}{\zeta_{{2}}}^{2}{\omega}^{2}{L}^{2}+2
\,v{k}^{4}{\rho_{{1}}}^{2}{\lambda_{B}}^{2}\rho_{{2}}{f}^{2}{L}^{2}+2\,v{k}^{6
}{\rho_{{1}}}^{2}\lambda_{B}\,{f}^{2}{L}^{2}
 \nonumber \\ && +{\zeta_{{1}}}^{2}{\omega}^{2}{
\lambda_{B}}^{2}{\rho_{{2}}}^{2}{L}^{2}+{v}^{2}{k}^{4}{\rho_{{1}}}^{2}{\zeta_{
{2}}}^{2}{\omega}^{2}{L}^{2}+2\,{\zeta_{{1}}}^{2}{\omega}^{2}{k}^{2}
\lambda_{B}\,\rho_{{2}}{L}^{2} \nonumber \\
& & +2\,\zeta_{{1}}{\omega}^{2}{\lambda_{B}}^{2}\rho_{{2
}}{f}^{2}\rho_{{1}}\zeta_{{2}}{L}^{2}+2\,v{k}^{2}{\rho_{{1}}}^{2}{\zeta_{{
2}}}^{2}{\omega}^{2}\lambda_{B}\,{f}^{2}{L}^{2}+2\,{v}^{2}{k}^{6}{\rho_{{1}}}^
{2}\lambda_{B}\,\rho_{{2}}{L}^{2}\nonumber \\ &&
+{v}^{2}{k}^{4}{\rho_{{1}}}^{2}{\lambda_{B}}^{2}{
\rho_{{2}}}^{2}{L}^{2}+{v}^{2}{k}^{8}{\rho_{{1}}}^{2}{L}^{2}+2\,{k}^{2}{
\zeta_{{2}}}^{2}{\omega}^{2}\lambda_{B}\,{f}^{2}\rho_{{1}}L+2\,{k}^{4}{\lambda_{B}
}^{2}\rho_{{2}}{f}^{2}\rho_{{1}}L
\nonumber \\ & &  +{k}^{4}{\zeta_{{2}}}^{2}{\omega}^{2}+2\,
{k}^{6}\lambda_{B}\,\rho_{{2}}+{k}^{4}{\lambda_{B}}^{2}{\rho_{{2}}}^{2}+2\,L{k}^{4
}{\zeta_{{2}}}^{2}{\omega}^{2}v\rho_{{1}}\nonumber \\ & &\left. +2\,L{k}^{4}{\lambda_{B}}^{2}{\rho_{{
2}}}^{2}v\rho_{{1}}+4\,L{k}^{6}v\rho_{{1}}\lambda_{B}\,\rho_{{2}}
\right).
\end{eqnarray}
Similar results hold for the cases of the other structure factors of the
system. 
This is the structure factor of the polymers in the coupled systems of ions and
counterions, where there is an excluded volume interaction between the polymer
segments, and a dynamic effective interaction, and dynamic alteration to the
distribution of the polyions.  When the limits $\rho_{2}\rightarrow 0$,
$\lambda_{B} \rightarrow 0$ are taken, the expression becomes equivalent to that
of Fredrickson and Helfand.\cite{FredricksonHelfand}  

Although this factor
looks rather complicated,
the form of the equation can be simplified significantly by noting the
definition of the Debye parameter,
\begin{equation}
\kappa^{2} = \lambda_{B} \rho_{2}
\end{equation}
and by introducing the quantity
\begin{equation}
{\tilde \xi}^{-2} = \rho_{1} v
\end{equation}
which differs from the conventional definition of the screening parameter by
Edwards\cite{EdwardsScreeningLength} by a factor of 4.  

A plot of this structure factor is shown in Figure 1, where convenient values
have been used for the parameters.  The polyelectrolyte peak is clearly
visible and moves to greater values of $k$ with increasing frequency.  The
peak has already been discussed in the context of statistical mechanics by
various authors.\cite{VilgisBorsali}

We find that broadening together with heightening in both $k$ and $\omega$ directions of the
polyelectrolyte peak is a result of either a decrease of the polyion charge
density or of an increase of Debye parameter.

\subsection{Diffusional behaviour}

By inspection of the fraction, equation 3.2,  it is easy to identify that
the numerator is a second order polynomial in $\omega$ and the denominator,
when expanded is a polynomial of fourth order in the frequency.  This suggests
that the diffusion in the present approximation has two modes, which can be
extracted by breaking the expression 3.2 into its partial
fractions.
The result of this manipulation is:
\begin{equation}
G_{11} = \frac{A_{1}}{\Gamma_{1} + \omega^{2}} + \frac{A_{2}}{\Gamma_{2} + \omega^{2}},
\end{equation}
where the modes associated with the amplitudes $A_{1}$ and
$A_{2}$ have the following values which have been expanded to their
lowest orders in $k^{2}$:
\begin{eqnarray}
\sqrt{\Gamma_{1}}/k^{2} & \simeq & \frac{\rho_{1}f^{2} +\rho_{1}\rho_{2}v +
  L^{{-1}}\rho_{2}    }{\zeta_{1}\rho_{2}  + f^{2} \rho_{1}\zeta_{2}} = D_{c}
\\
\sqrt{\Gamma_{2}} & \simeq & \frac{\lambda_{B}}{\zeta_{1}\zeta_{2}}\left( f^{2}
  \rho_{1} \zeta_{2} +\zeta_{{1}} \rho_{2} \right). 
\end{eqnarray}

This behavior to is to be anticipated in a system where there are two
components coupled by an interaction such as the Coulomb interaction, and has
been previously deduced within the context of a far simpler assumption about
the dynamics of the polyelectrolyte system.
 The mode
$\Gamma_{2}$ has already previously been identified as the plasmon mode.  It
is not proportional to $k^{2}$ at large length-scales and agrees with that
found by other methods.\cite{VilgisBorsali}

The cooperative diffusion coefficient is given by the expression 3.6.
Here we see that this is also given as in the paper of Vilgis
and Borsali,\cite{VilgisBorsali} with the 
limiting proportionality $ v\rho_{1} + 1/L$.  This is the
expected behavior.  We note that in the present scheme of approximation 
the cooperative diffusion coefficient, $D_{c}$, depends upon the interaction
only through the linear charge density of the chains and the counterion
concentration, whereas the plasmon mode is directly proportional to the
Bjerrum length.


\section{Discussion}

In the first scheme of integration of section~II, where the
immediate integration over the fluid velocity field degrees of freedom leads
directly to the nonlinear term in the polymer density, which is the only term
coupling the nature of the interaction into the hydrodynamic description and
vice-versa.  Under the RPA for the transversal components too it is clear that
this coupling term plays no contribution whatsoever, and then the dynamics of
this system would be exactly as in the case derived by Fredrickson and
Helfand.\cite{FredricksonHelfand}   
For frequencies $\omega L^{2}/\pi^{2}\zeta_{1} \gg 1$ and at large length
scales
$kR_{g} \ll 1$ a macroscopic solution shear viscosity is found $\eta' =
\eta ( 1 + \rho_{1} [\eta]_{R})$, with the intrinsic viscosity of the Rouse model
$[\eta]_{R}= L\zeta_{1}12\eta$.\cite{FredricksonHelfand}  At length scales
below the radius of gyration, Fredrickson and Helfand find a hydrodynamic
screening length $\xi^{2} \propto \eta/\rho_{1}\zeta_{1}$, which screens the
hydrodynamics such that at the smallest distances the viscosity is again
$\eta$.
This is completely {\em independent} of the nature of the interaction between
the polymer segments, as it contains only terms due to the connectivity of the
polyion chains.

However, the dynamic scattering intensity  derived above can be implemented in an
alternative computation of the reduced viscosity in the Rouse as well as the
Zimm cases by making use of the results of Hess' and Klein's mode-mode
coupling computations,\cite{HessKlein} as has been done by Vilgis, Borsali,
and Benmouna,\cite{VB1,VB2} by implementation of the formula $
\eta_{r}$ $ = \frac{k_{B}T \rho_{1}^{2}}{16\pi^{3}\eta c_{p}} $ $\int_{0}^{\infty}dt
\int_{0}^{\infty} dk\, S^{2}({\bf k},t) \frac{k^{2}}{S^{4}(k))}
\frac{\partial H(k)}{\partial {\bf k}}$.
The function $H(k)$ represents the particle-particle correlation function.
Hitherto this formula could only be implemented by estimating the first
cumulant frequency for the dynamical scattering intensity, but here we have
already computed this from the more complete dynamical formalism.  We have
also shown that in the appropriate limits, the behavior of our more
complicated system is like that derived by more conventional means, so we
expect the occurrence of a viscosity peak to be qualitatively retained in this
formalism.


It should be pointed out that the assumption of homogeneity of the separate
constituents of charge and species in the solution, that these are in no
way correlated when the Gaussian approximation of the collective coordinates
is made, provides the approximation which leads to the Debye-H\"uckel
potential.  This can also be derived in the other approximations, such as the
linearization of the Poisson-Boltzmann equation.  The RPA is valid for
systems in which the temperature is high such that the various counterions
can be understood to move almost independently and, and when the polymer
density is high, such that the chains have a Gaussian end-to-end distribution.
The success of this calculation has been tested in many and varied fields.
As has been shown the RPA can be applied to the dynamical case too, with the
complicating factor of the occurrence of additional fields;  however, these
are also quadratic.

In polyelectrolytes the effect of the condensation of counterions
on the chains has been known for some time (see refs 31, 32  and references
therein).  The critical onset of
this condensation for monovalent ions which is derived from models
of charged cylindrical objects occurs in Manning's theory\cite{Man1}
when $\lambda_B f \geq 1$ (and at a lower clustering temperature
in the case of solutions of rod-like polyelectrolytes; see Levin\cite{Levin} and
Levin and Barbosa.\cite{LevinBarbosa})
such that we consider here the case of high temperature and low charge
density.  Indeed, this other regime of counterion condensation on flexible 
objects is less well
understood and would necessitate a different treatment than the RPA. 

Firstly, it is noticed that the presence of the hydrodynamic interactions
generates a  term to third order in the density of the fields.  Attempts to
approximate this as a linear or quadratic contribution to the integrand are
bound to fail.  This is because of the decoupling in the RPA of the fields
relevant to the transverse and parallel ${\bf k}$ directions.  The cubic term
is responsible for the coupling of the hydrodynamic properties of the system
to the remaining dynamic properties, and we can see how its neglect, makes the
diffusive motion not feature any properties of the solvent, except through the
definition of the friction parameter, 
$\zeta = 6\pi\eta a$,
where $a$ represents the particle radius.  
In order to understand more fully the role of the coupling (not only that of the
connectivity) but especially of the nature of the interaction into the
hydrodynamic properties, it becomes necessary to use an altered approach to
the dynamics.  This falls beyond the scope of this paper, and will be
discussed in a subsequent publication.\cite{MV2}

We have given the dynamical structure factor of the system under discussion,
and shown that it reproduces the polyelectrolyte peak at sufficiently low
frequencies.  We have also discussed the diffusive behavior of the system, and
reconfirmed the occurrence of the ``plasmon'' mode using the dynamical formalism.

\section*{Acknowledgment}

The funding of this work by the Deutsche Forschungsgemeinschaft:
Schwerpunkt Polyelektrolyte is most gratefully acknowledged.


\section*{Appendix: RPA Results}

The RPA has been computed before by Fredrickson and
Helfand.\cite{FredricksonHelfand}  Here we present the relevant
results for the case
$kR_{g} \ll 1$.
\begin{eqnarray}
\left\langle \rho_{1}(-{\bf k},-\omega) \rho_{1}({\bf k}, \omega) \right\rangle_{\rm RPA} & = 
& \frac{2k^{2}\zeta_{1}}{\omega^{2}\zeta_{1}^{2} + k^{4}L^{-2}} \\
\left\langle \rho_{2}(-{\bf k},-\omega) \rho_{2}({\bf k}, \omega) \right\rangle_{\rm RPA} & = 
& \frac{2k^{2}\zeta_{2}}{\omega^{2}\zeta_{2}^{2} + k^{4}} \\
\left\langle \rho_{1}(-{\bf k},-\omega) {\bf k}\cdot
{\boldsymbol{\sigma}}_{1}({\bf k}, \omega) \right\rangle_{\rm RPA} & = 
& \frac{2k^{2}}{\omega\zeta_{1} + ik^{2}L^{-1}} \left( \frac{\zeta_{1}}{i}\right) \\
\left\langle \rho_{2}(-{\bf k},-\omega) {\bf k}\cdot
{\boldsymbol{\sigma}}_{2}({\bf k}, \omega) \right\rangle_{\rm RPA} & = 
& \frac{2k^{2}}{\omega\zeta_{2} + ik^{2}}\left({\frac{\zeta_{2}}{{i}}}\right).
\end{eqnarray}
Apart from the nontrivial remaining averages $\left\langle \pi_{\alpha} \pi_{\alpha}\right\rangle$ and
$\left\langle \pi_{\alpha} ({\bf 1} - {\bf \hat{k}\hat{k}}) \sigma_{\alpha}\right\rangle$ all remaining
pairwise correlations are zero.




\begin{figure}
\epsffile{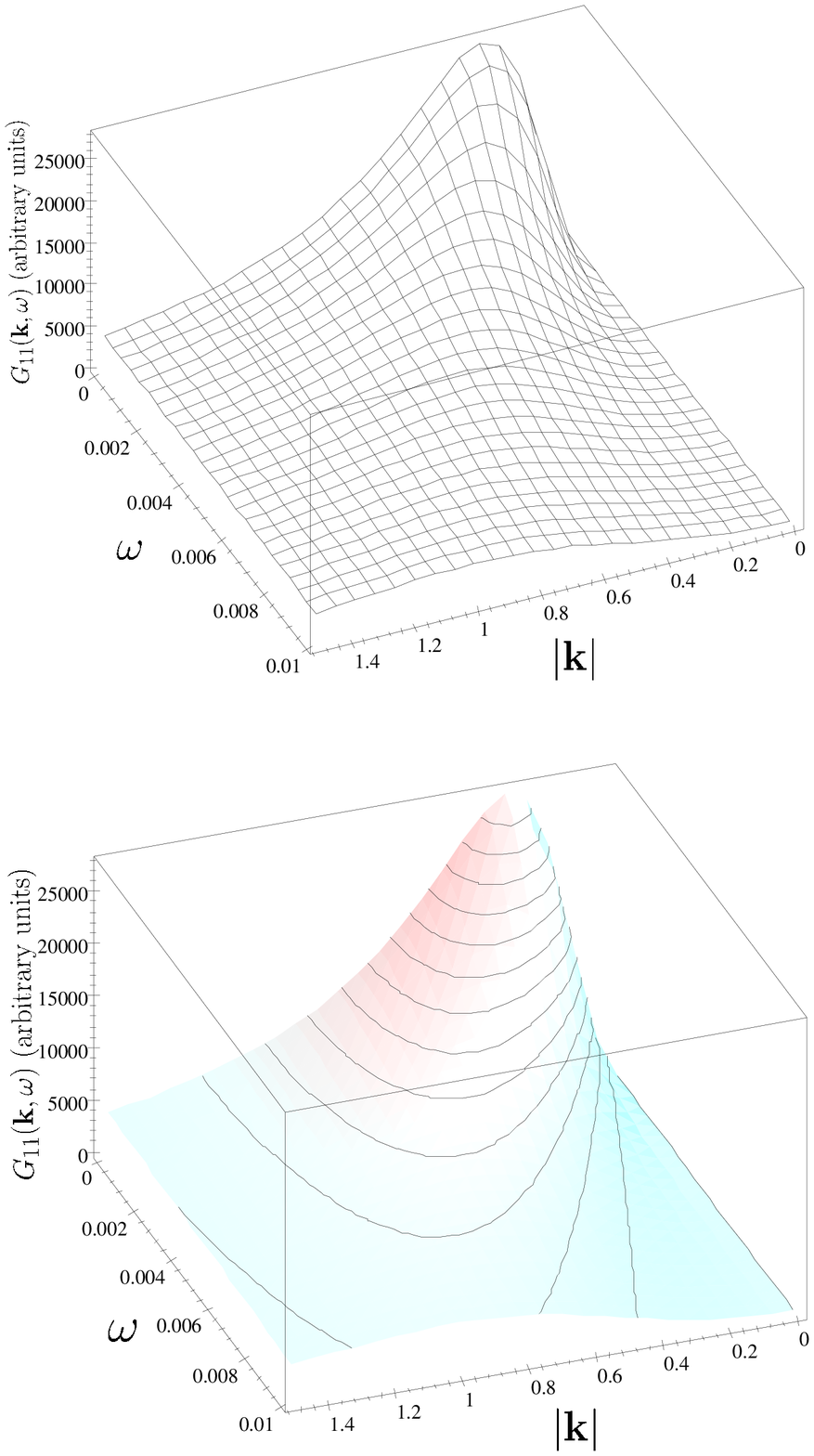}
\caption{The dynamical structure factor for $\zeta_{1}=\zeta_{2}=1$,
  $\rho_{1}=1$,
$f=0.03$, $L=100$, $v=0$, $\kappa=0.01$, and $\lambda_{B}=1$.
The graphs show clearly how the peak in the dynamic structure
factor broadens in the $\omega$--direction.  The maximum width in $\omega$
occurs at a momentum transfer vector larger than the location of the
scattering peak.
}
\end{figure}

\end{document}